\documentclass[preprint]{elsarticle}
\usepackage[utf8]{inputenc}
\usepackage{epstopdf}
\usepackage{color}
\usepackage{amsmath}
\usepackage{units}
\usepackage{rotating}
\begin{document}

\title{Quantum Correlations in Deutsch-Jozsa Algorithm via Deterministic Quantum Computation \\with One Qubit Model}

\author[a1,a2]{Márcio M. Santos\corref{cor1}}
\ead{marcio.santos@ufvjm.edu.br}
\author[a3]{Eduardo I. Duzzioni}
\ead{duzzioni@gmail.com}
%\author[els]{S.~Pepping\corref{cor2}\fnref{fn1,fn3}}
%\ead[url]{http://www.elsevier.com}

\cortext[cor1]{Corresponding author}
\address[a1]{Instituto de Ciência, Engenharia e Tecnologia, Universidade Federal dos Vales do Jequitinhonha e Mucuri\\ Rua do Cruzeiro 01, Jardim São Paulo, 39803-371\\
Teófilo Otoni, Minas Gerais, Brazil\\
}
%{River Valley Technologies, SJP Building,
%Cotton Hills, Trivandrum, Kerala, India 695014}
\address[a2]{Instituto de Física, Universidade Federal de Uberlândia\\ Av. João Naves de Ávila 2121, Santa Mônica, 38400-902\\
Uberlândia, Minas Gerais, Brazil\\}
%{River Valley Technologies, 9, Browns Court,
%Kennford, Exeter, United Kingdom}
\address[a3]{Departamento de Física, Universidade Federal de Santa Catarina\\ Campus Universitário Reitor João David Ferreira Lima, Trindade, 88040-900 \\ Florianópolis, Santa Catarina, Brazil\\}

\begin{abstract}
Quantum correlations have been pointed out as the most likely source of the speed-up in quantum computation. Here we analyzed the presence of quantum correlations in the implementation of Deutsch-Jozsa algorithm running in the DQC1 and DQCp models of quantum computing. For some balanced functions, the qubits in DQC1 model are quantum correlated just in the intermediate steps of the algorithm for a given decomposition into one and two qubits gates. In the DQCp model the final state is strongly quantum correlated for some balanced functions, so that the pairwise entanglement between blocks scales with the system size. Although the Deutsch-Jozsa algorithm is efficiently implemented in both models of computation, the presence of quantum correlations is not a sufficient property for computational gain in this case, since the performance of the classical probabilistic algorithm is better than the quantum ones. The measurement of other qubits than the control one showed to be inefficient to turn the algorithm deterministic.
\end{abstract}


\maketitle

\section{Introduction}	%) A SECTION HEADING

Quantum correlations have been pointed out as a resource for quantum computation. To observe advantage of pure state quantum computation over classical computing entanglement is seen as a necessary resource\cite{Jozsa,Nest2012}. However, such resource does not seem to be so essential for quantum computation with mixed states. It was observed that the amount of entanglement present in the trace evaluation of a unitary matrix realized in the Deterministic Quantum Computation with one qubit (DQC1) model could not explain the resulting speed-up \cite{entan DQC1}. Other examples using de DQC1 model which present quantum advantage are the Shor\textquoteright s factorization algorithm \cite{Parker}, the measurement of the average fidelity decay of a quantum map \cite{Poulin}, and the approximation of the Jones Polynomial \cite{Shor2008}. This model of computation have already been implemented in an optical system \cite{Lanyon} and in Nuclear Magnetic Resonance \cite{Ryan,Marx}. 

\begin{figure}
\begin{center}
\includegraphics[clip,scale=0.2]{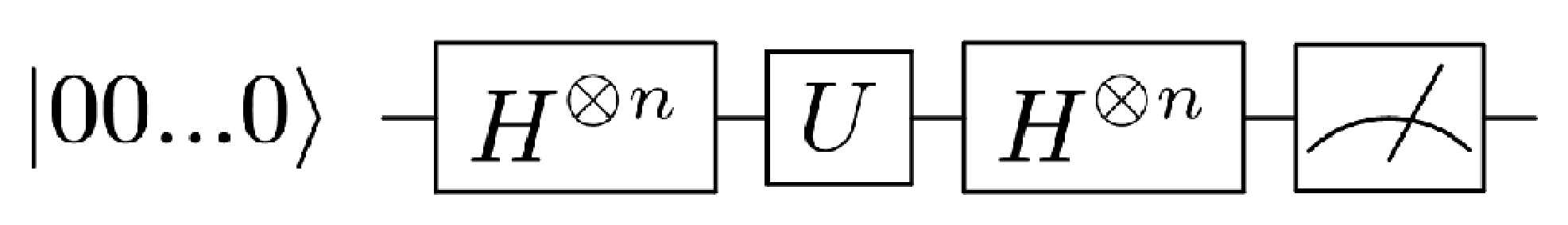}\protect\caption{\label{fig:FigCollins}Collins version for the Deutsch-Jozsa algorithm. $H$ is the Hadamard gate and $U$ is the unitary matrix encoding the function $f$.}
\end{center}
\end{figure}

The first quantum algorithm, introduced by D. Deutsch in 1985 \cite{Deutsch1985}, aimed to decide if a function $f:\{0,1\}\rightarrow\{0,1\}$ is constant or balanced. Although the first version of this algorithm was probabilistic, improvements on it showed that it is possible to know with certainty the function class with just one measurement \cite{Mosca1997}, while in the classical case two evaluations of $f$ are necessary. The generalization of Deutsch algorithm to an input of $n$ qubits was made by D. Deutsch and R. Jozsa in 1992 \cite{DeutschJozsa}. In this case the function $f:\{0,1\}^{n}\rightarrow\{0,1\}$ is said to be constant if $f(j)=0$ or $f(j)=1$ for all $j$ $(j=0,...,2{}^{n}-1)$ and balanced if $f(j)=0$ for half of the $j$ values and $f(j)=1$ for the remaining $j$ values. Classically, it will take $2$ to $2^{n-1}+1$ evaluations of $f$ to know the function class with certainty. In quantum computation the Deutsch-Jozsa algorithm gives the exactly answer to the problem with just $n$ individual qubit measurements. In Ref. \cite{Mosca1997} the authors use $n+1$ qubits to solve this problem, while the Collins version of this algorithm uses only $n$ qubits \cite{DJ Collins}. The Collins version is represented by the circuit in Fig. \ref{fig:FigCollins}, where the unitary $U$ encodes the function $f$. After the $n$ first Hadamard gates the register is in an equal superposition of $2{}^{n}$ states $\left|j\right\rangle $. The action of $U$ on $\left|j\right\rangle $ is $U\left|j\right\rangle =(-1)^{f(j)}\left|j\right\rangle $, which means $U$ is a matrix of the form 
\begin{equation}
U=\sum_{j}(-1)^{f(j)}\left|j\right\rangle \left\langle j\right|.\label{eq:TU-1}
\end{equation}
Applying again the $n$ Hadamard gates the readout of the algorithm can be made by projecting the final state on the state $\left|j=0\right\rangle $, giving the result 
\begin{equation}
\frac{1}{2^{n}}\sum_{j=0}^{2^{n}-1}(-1)^{f(j)}=\left\{ \begin{array}{c}
\pm1\quad\textrm{if \ensuremath{f}\:\ is constant,}\\
0\quad\textrm{if \ensuremath{f}\:\ is balanced.}
\end{array}\right.\label{eq:deutsch-1}
\end{equation}

\begin{figure}[h]
\begin{center}
\includegraphics[clip,scale=0.3]{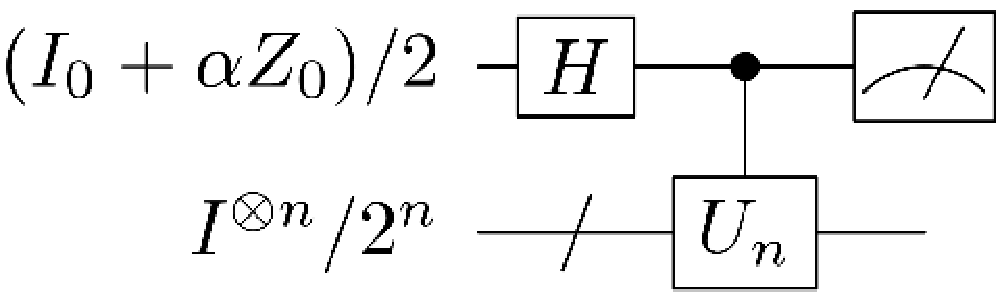}\protect\caption{\label{fig:DQC1-circuit}DQC1 circuit used to evaluate the normalized trace of a unitary matrix $U_{n}$. A qubit is prepared in a state with some degree of purity (top rail) and a set of $ n $ qubits is prepared in the maximally mixed state (bottom rail). A Hadamard gate is applied on the top qubit before the application of the controlled $U_{n}$. To finish, a measurement is made upon the top qubit.}
\end{center}
\end{figure}

The structure of $U$ in Eq. (\ref{eq:TU-1}) allows the definition of the function class of $f$ by the evaluation of its normalized trace. An efficient way to evaluate the trace of a unitary matrix is given by the DQC1 model \cite{DQC1}. The DQC1 model is composed by $n+1$ qubits, where $n$ qubits are in the fully mixed state $I{}^{\otimes n}/2^{n}$ and only one qubit presents a degree of purity controlled by $\alpha\;(0<\mathrm{\alpha<1})$, as represented by the circuit in Fig. \ref{fig:DQC1-circuit}. The system initial state is $\rho_{I}=2^{-(n+1)}(I_{0}+\alpha Z_{0})\otimes I{}^{\otimes n}$, where the index $0$ refers to the semi-pure qubit (control qubit), $I$ is the identity matrix, and $Z$ is the Pauli matrix $\sigma_{Z}$. 

Just before the measurement the state of the system is
\begin{equation}
\rho=\frac{1}{2^{n+1}}\left(I_{0}\otimes I_{n}+\alpha X_{0}\otimes U_{n}\right),\label{eq:finalstate}
\end{equation}
where $X_{0}$ is the Pauli matrix $\sigma_{x}$. One way of implementing the Deutsch-Jozsa algorithm via DQC1 model is choosing the unitary matrix $U_{n}$ as $U$ defined in Eq. (\ref{eq:TU-1}) \cite{Arvind}. In this case, the state of the control qubit is characterized by $\left\langle \sigma_{x}\right\rangle =\alpha\frac{1}{2^{n}}\sum_{j=0}^{2^{n}-1}(-1)^{f(j)}$ and $\left\langle \sigma_{y}\right\rangle =\left\langle \sigma_{z}\right\rangle =0$. Thus, if a $\sigma_{x}$ measurement is performed on this qubit, the result for its expected value is $\left\langle \sigma_{x}\right\rangle =0$ and variance $\triangle\sigma_{x}=1$ for a balanced function and $\left\langle \sigma_{x}\right\rangle =\pm\alpha$ with variance $\triangle\sigma_{x}=\sqrt{1-\alpha^{2}}$ for a constant function. 

The Deutsch-Jozsa algorithm was approached by different models of computation: probabilistic classical computation \cite{Arvind,preskill}, circuital quantum computation with pure states\cite{Mosca1997,DJ Collins}, ensemble quantum computation \cite{Arvind}, adiabatic quantum computation \cite{Das,Wei}, one-way quantum computation \cite{Chaves2011}, dissipative quantum computation \cite{Santos2012}, and blind quantum computation \cite{Barz2012}.

In this work we study the presence of quantum correlations in the Deutsch-Jozsa algorithm performed in the DQC1 and DQCp models of quantum computation. In the next section we develop a circuit to perform the computation composed by a universal set of gates. Then, considering the DQC1 model we evaluate the presence of quantum correlations after the application of each quantum gate. To observe the generation of correlations with an initial pure state we evaluate, in section 3, the negativity generated by the realization of the same algorithm performed in the DQCp model. We then present a discussion followed by our conclusions.

\section{Quantum Correlations in Deutsch-Jozsa Algorithm via DQC1 Model}

Now we analyze the correlations present in the Deutsch-Jozsa algorithm implemented via DQC1 model. The final state of the computation (\ref{eq:finalstate}) can be written as 
\begin{align}
\rho & =\sum_{j=0}^{2^{n}-1}(1/2^{n+1})\left[\left|0\right\rangle \left\langle 0\right|+\alpha(-1)^{f(j)}\left|0\right\rangle \left\langle 1\right|\right.\nonumber \\
 & \quad\quad\left.+\alpha(-1)^{f(j)}\left|1\right\rangle \left\langle 0\right|+\left|1\right\rangle \left\langle 1\right|\right]\otimes\left|j\right\rangle \left\langle j\right|\nonumber \\
 & =\sum_{j=0}^{2^{n}-1}(1/2^{n+1})(\left|a_{j}\right\rangle \left\langle a_{j}\right|+\left|b_{j}\right\rangle \left\langle b_{j}\right|)\otimes\left|j\right\rangle \left\langle j\right|,\label{eq:statef}
\end{align}
where $\left|a_{j}\right\rangle =cos\phi\left|0\right\rangle +(-1)^{f(j)}sin\phi\left|1\right\rangle $, $\left|b_{j}\right\rangle =sin\phi\left|0\right\rangle +(-1)^{f(j)}cos\phi\left|1\right\rangle $, and $sin(2\phi)=\alpha$ \cite{Datta Thesis}. Particularly for $\alpha=1$ the final state is 

\begin{equation}
\rho=\sum_{j=0}^{2^{n}-1}(1/2^{n})\left|f(j)\right\rangle \left\langle f(j)\right|\otimes\left|j\right\rangle \left\langle j\right|,\label{eq:statef1}
\end{equation}
with $\left|f(j)\right\rangle =(\left|0\right\rangle +(-1)^{f(j)}\left|1\right\rangle )/\sqrt{2}$. 

It is easy to observe from Eqs. (\ref{eq:statef}) and (\ref{eq:statef1}) that the state $\rho$ is separable for any partition, since the $|j\rangle$ states describe the computational basis. We remember that any bipartite separable state can be cast under three categories: \emph{i)} classical-classical (CC) states with the form $\rho=\sum_{i}p_{i}\left|i_{A}\right\rangle \left\langle i_{A}\right|\otimes\left|i_{B}\right\rangle \left\langle i_{B}\right|$ where $\left\{ \left|i_{A(B)}\right\rangle \right\} $ is an orthonormal set and $\{p_{i}\}$ is a probability distribution, \emph{ii)} classical-quantum (CQ) states with the form $\rho=\sum_{i}p_{i}\left|i_{A}\right\rangle \left\langle i_{A}\right|\otimes\rho_{i}^{B}$ where $\left\{ \rho_{i}^{B}\right\} $ are quantum states, and \emph{iii)} fully quantum states (QQ) with the form $\rho=\sum_{i}p_{i}\rho_{i}^{A}\otimes\rho_{i}^{B}$\cite{Piani}. Rewriting the state in Eq. (\ref{eq:statef}) in the form
\begin{equation}
\sum_{\substack{i = +,-\\  0<j<2^n - 1}}p_{i,j}\left|i\right\rangle \left\langle i\right|\otimes\left|j\right\rangle \left\langle j\right|,\label{eq:statef1b}
\end{equation}
where $ \left|\pm\right\rangle = (\left|0\right\rangle \pm \left|1\right\rangle)/\sqrt{2}$ and $p_{\pm,j} = (1\pm\alpha (-1)^{f(j)})/2$, it becomes clear that the final state of the computation is a CC state. Therefore $\rho$ has no quantum correlations. Indeed, any quantum discord-like measure over any bipartition should confirm this statement \cite{Ollivier2001,Henderson2001,Modi2012,Rulli}.

After performing $\sigma_{x}$ measurements on the control qubit the best scenario to discriminate between constant and balanced functions occurs when $\alpha=1$. The Deutsch-Jozsa algorithm is implemented efficiently via DQC1 model, since the expected value of $\sigma_{x}$ must be known with a given precision, which is independent of the number $n$ of mixed qubits. In Ref. 15
%\cite{Arvind } 
the authors show that this quantum algorithm (for $\alpha=1$) has at most a good performance as the classical probabilistic algorithm. Therefore, the quantum and classical versions of the Deutsch-Jozsa algorithm discussed here have equivalent performance. Such result is not obvious, because it is possible that quantum correlations are present in intermediate states of the computation even when the initial and final states do not have any. Although this is not the case, quantum correlations may be related to the speedup of quantum computation since the quantum computer can evolve through states that use a smaller number of gates in the quantum solution \cite{Datta correlations middle}.

\begin{center}

\begin{sidewaysfigure}
%\rule{0.75\textheight}{0.5\textheight}
\textcolor{white}{\rule{1cm}{4cm}}
%\textcolor{white}{\rule{6cm}{6cm}}‎
\includegraphics[scale=0.5]{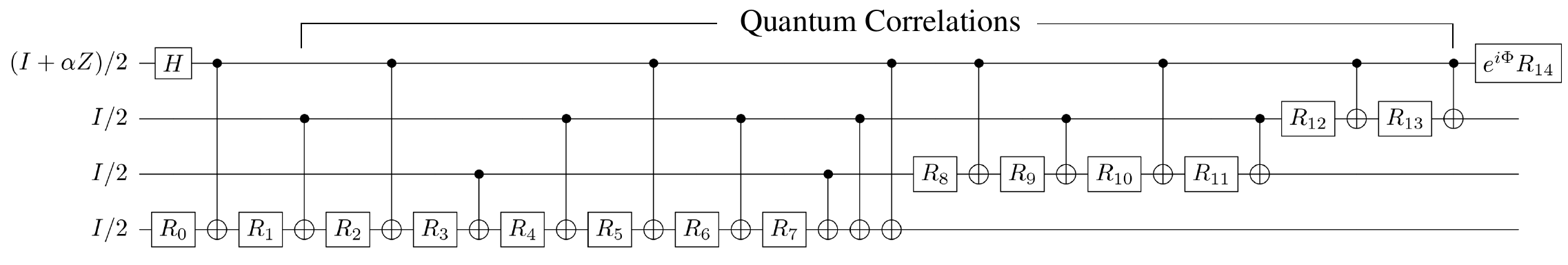}\protect\caption{\label{fig:Synthesized-Deutsch-Jozsa-algori-1}Synthesized Deutsch-Jozsa algorithm implemented via DQC1 model for three mixed qubits. Depending on the balanced function, the qubits can be quantum correlated along the steps highlighted on the top of the figure. Here, the operator $R_{j}\equiv R_{z}^{l}(\theta_{j})=e^{-i\theta_{j}/2}\left|0\right\rangle _{l}\left\langle 0\right|+e^{i\theta_{j}/2}\left|1\right\rangle _{l}\left\langle 1\right|$ rotates the state of the $l$-th qubit by an amount $\theta_{j}$ around the $z$ axis. The angles of rotation are defined by $\theta_{0}=-\theta_{1}\equiv-\pi(f_{0}-f_{1}+f_{2}-f_{3}+f_{4}-f_{5}+f_{6}-f_{7})/8$, $\theta_{2}=-\theta_{3}\equiv\pi(f_{0}-f_{1}+f_{2}-f_{3}-f_{4}+f_{5}-f_{6}+f_{7})/8$, $\theta_{4}=-\theta_{7}\equiv-\pi(f_{0}-f_{1}-f_{2}+f_{3}-f_{4}+f_{5}+f_{6}-f_{7})/8$, $\theta_{5}=-\theta_{6}\equiv-\pi(f_{0}-f_{1}-f_{2}+f_{3}+f_{4}-f_{5}-f_{6}+f_{7})/8$, $\theta_{8}=-\theta_{9}\equiv-\pi(f_{0}+f_{1}-f_{2}-f_{3}+f_{4}+f_{5}-f_{6}-f_{7})/8$, $\theta_{10}=-\theta_{11}\equiv\pi(f_{0}+f_{1}-f_{2}-f_{3}-f_{4}-f_{5}+f_{6}+f_{7})/8$, $\theta_{12}=-\theta_{13}\equiv-\pi(f_{0}+f_{1}+f_{2}+f_{3}-f_{4}-f_{5}-f_{6}-f_{7})/8$, and $\theta_{14}=2\Phi\equiv\pi(f_{0}+f_{1}+f_{2}+f_{3}+f_{4}+f_{5}+f_{6}+f_{7})/8.$}

\end{sidewaysfigure}
\end{center}

To investigate the birth and death of quantum correlations in the implementation of the Deutsch-Jozsa algorithm, we use the procedure presented by S. Bullock and L. Markov to decompose a diagonal unitary operator in a sequence of one qubit rotations and CNOTs \cite{Bullock}. Such synthesis is general, so that it can describe any unitary applied over the Deutsch-Jozsa algorithm for balanced or constant functions. The decomposition was done for the two and three mixed qubits cases, where the later is presented in Fig. \ref{fig:Synthesized-Deutsch-Jozsa-algori-1}. The presence or absence of quantum correlations after each quantum gate in the synthesized algorithm is pointed out after writing the system state as a CC state or not. See the Appendix for details. For two mixed qubits case, corresponding to four values for the index $j$ ($j=00,01,10,11$), no quantum correlations are found in any step of the algorithm. In our decomposition for three mixed qubits, quantum correlations are found between the second and the last but one CNOT gates for some balanced functions. In this last case we found that there is no entanglement as measured by negativity (see the definition in Eq.(\ref{eq:neg})) \cite{negativity} evaluated for all steps in the synthesized algorithm considering different splits for all kind of functions: $i)$ a splitting that separates the control qubit from all the others and $ii)$ another one that puts the top two qubits in one partition and the bottom two in the other partition. We observe that the rotation angles $\theta_{j}$ present in the synthesis of the algorithm may have, among other values, $\pm\pi/4$ for some balanced functions. In these situations the operator $R_{j}$ becomes the $T$ (or $\pi/8$) gate, a unitary that lies outside the Clifford group. Despite of Gottesman-Knill\cite{GK} theorem and Bryan Eastin result (that a concordant computation can be simulated using a classical computer)\cite{Eastin} do not apply to these decompositions, the algorithm presented above can efficiently be simulated in a classical computer. 

For pure states quantum discord is equal to entanglement entropy, i.e., it measures entanglement between two parties \cite{Henderson2001}. In correspondence, Collins, Kim and, Holton arrived at a similar conclusion for the Deutsch-Jozsa algorithm implemented through the conventional pure state quantum computation model \cite{DJ Collins}. They found that no entanglement is generated between two qubits, while for three or more qubits some balanced functions generate entanglement among them. Chaves and de Melo showed that there are functions for which it is possible to implement the Deutsch-Jozsa algorithm in the one-way quantum computation method with decoherence from a state that presents only classical correlations \cite{Chaves2011}. Arvind, Dorai and, Kulmar implemented the Deutsch-Jozsa algorithm in a NMR experiment and observed the absence of entanglement for the one and two qubits cases and entanglement generation for some balanced functions in the three qubit case \cite{Dorai}. By using pure state quantum computation, Kenigsberg, Mor, and Ratsaby found the maximal sub-problem that can be solved without entanglement \cite{Kenigsberg}.

\section{Quantum Correlations in Deutsch-Jozsa Algorithm via DQCp model}

The basic idea of the deterministic quantum computation with pure states (DQCp model) is to reproduce in the answer qubit the expectation values of DQC1 model for the control qubit \cite{DQC1}. The same results for the Deutsch-Jozsa algorithm in DQC1 model presented above, that is, the expected values and variances of $\sigma_{x}$ for the control qubit, can be achieved if $\alpha=1$ and the $n$ mixed qubits in the DQC1 circuit are initialized in the state $\left|+\right\rangle ^{\otimes n}=\left[(\left|0\right\rangle +\left|1\right\rangle )/\sqrt{2}\right]^{\otimes n}$. This particular result, by its turn, demonstrates that to solve oracle problems in a quantum computer running via DQC1 model can be as powerful as in a quantum computer running via DQCp model \cite{DQC1}. An important difference between DQC1 and DQCp models is that with pure initial states the circuit can generate significant amounts of entanglement among the qubits at the end of the computation, while with mixed states none quantum correlation can be generated. 

To verify this hypothesis, we quantify entanglement among different partitions through negativity, a measure which has the advantage of being easily evaluated for a general bipartite mixed state \cite{negativity}. Its expression is given by

\textcolor{black}{
\begin{equation}
\mathcal{N}\left(\rho\right)=\frac{\left\Vert \rho^{T_{A}}\right\Vert _{1}-1}{2},\label{eq:neg}
\end{equation}
where $\left\Vert O\right\Vert _{1}=\textrm{tr}\sqrt{O^{\dagger}O}$ is the trace norm of operator $O$ and the partial transposition on subsystem $A$ is denoted by $\rho^{T_{A}}$ (it could have also been defined with partial transposition on subsystem $B$). The range of values for negativity goes from zero to $(d-1)/2$, where $d$ is the smaller partition between $A$ and $B$.}

\begin{figure}[h]
\begin{center}
\includegraphics[clip,scale=0.3]{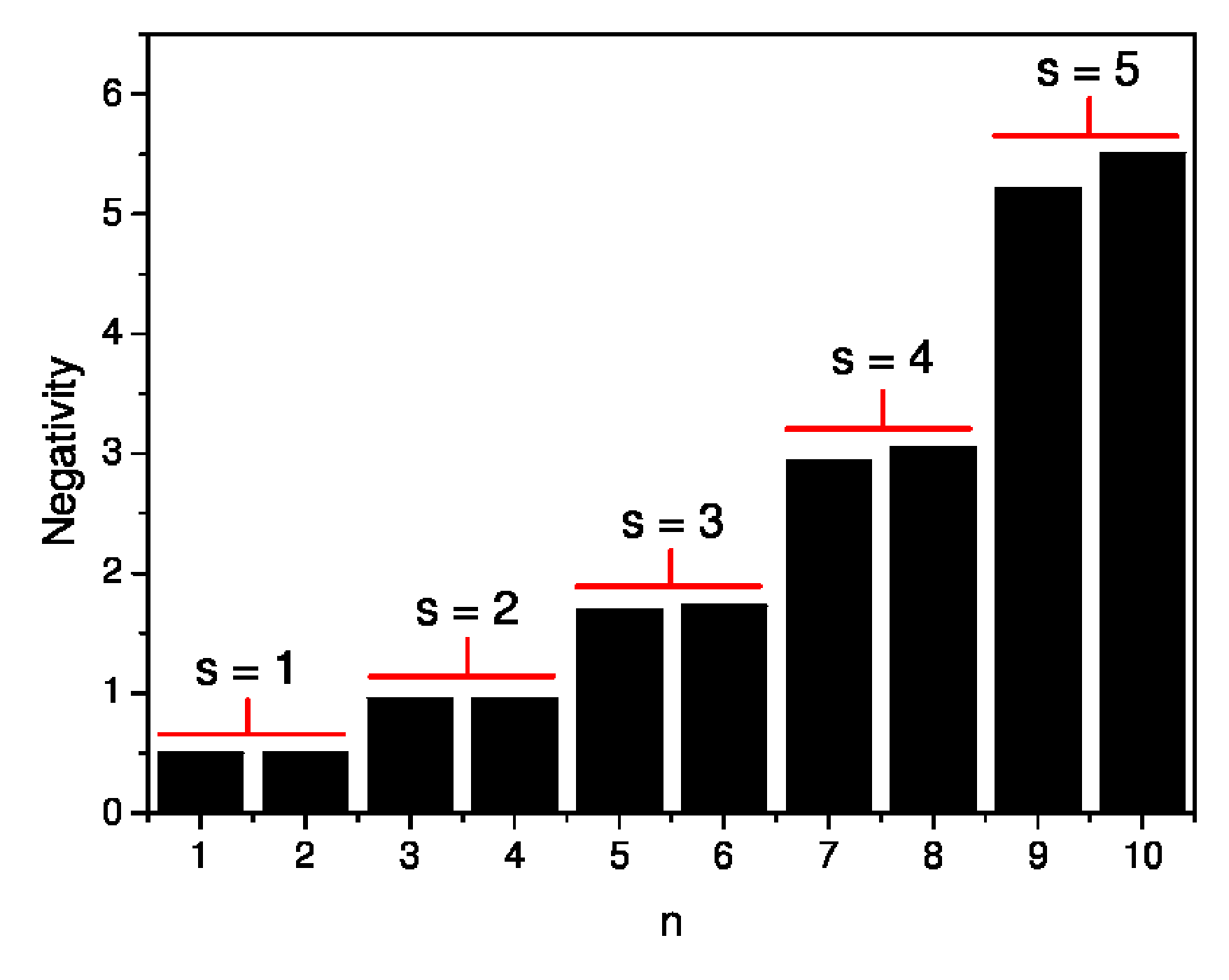}

\protect\caption{Negativity for the final state in the realization of Deutsch-Jozsa algorithm via DQCp model as a function of the number of qubits $n$. The values are relative to the splitting separating the $(n+1)/2$ top qubits and the $(n+1)/2$ bottom qubits, for $n$ odd, and the $n/2$ top qubits and $n/2+1$ bottom qubits for $n$ even. The sequence of approximately constant values is due to the negativity limitation by the dimension of the smaller partition. Here, $s$ is the number of qubits in the smaller partition.\label{fig:Negativity}}
\end{center}
\end{figure}

The algorithm was run 50 times with random balanced functions for a number of work qubits from 1 to 10, and the negativity was evaluated for two different splits: $i)$ a split that separates the $(n+1)/2$ top qubits and the $(n+1)/2$ bottom qubits for $n$ odd, and $ii)$ the $n/2$ top qubits and $n/2+1$ bottom qubits for $n$ even. The maximum value of the negativity achieved for each number of qubits is shown in Fig. \ref{fig:Negativity}. The resulting curve presents an overall increasing pattern, with the characteristic of sequential values of the negativity being approximately constant since it is limited by the dimension of the smaller partition \cite{Datta Thesis}. Differently from the results presented above for DQC1 model, even for the case with $n=1$ the states of the system are quantum correlated, a behavior that remains as $n$ increases.

\section{Discussion }

The final states of the control qubit in the Deutsch-Jozsa algorithm implemented via DQC1 and DQCp models for the possible function classes do not have orthogonal support, which means they can not be distinguished with just one measurement as in the conventional pure state quantum computation \cite{Mosca1997,DJ Collins}. This makes such algorithm probabilistic. We tried to remove the probabilistic character of the solutions obtained through the models discussed above by using the DQC1\textbf{\large{}$_{k}$} model of computation \cite{Morimae2014}. In the later the the quantum circuit is similar to DQC1 circuit but $k$ qubits are measured instead of only one. Nevertheless it was unsuccessful. The same procedure was repeated for DQCp\textbf{\large{}$_{k}$} model of computation, but it was in vain again. A possible reason for these results relies on the state of the control qubit for balanced functions, which one is proportional to the identity operator. So it is impossible to distinguish it perfectly from any other state.

\section{Conclusions}

We reviewed the Deutsch-Jozsa algorithm implemented via DQC1 model and also expanded this idea to the DQCp model. In both models the initial state of the qubits is uncorrelated, quantum and classically. In the DQC1 model the final state of the algorithm does not possess any kind of correlations. Otherwise, in the DQCp model the qubits at the end of the algorithm are highly correlated for some balanced functions and the bipartite entanglement among the block of qubits increases with the system size. The Deutsch-Jozsa algorithm is efficiently implemented in these different models of computation. Independently of the existence or absence of quantum correlations among the qubits in each step of the algorithm, the quantum solution shows no advantage over that one given by the classical probabilistic algorithm. Some efforts in order to decide if the function class is constant or balanced by the realization of a $k$-collective measurement on the qubits in the DQC1\textbf{\large{}$_{k}$} and DQCp\textbf{\large{}$_{k}$} models of computation have been shown inefficient.

\section*{Acknowledgments}

The authors acknowledge the financial support from Brazilian agencies CAPES, FAPEMIG, CNPq, and Brazilian National Institute of Science and Technology for Quantum Information (INCT-IQ).

%\appendix
\section*{Appendix}

Here we show how to detect quantum correlations in the state of the system after every step of the synthesized algorithm for three mixed qubits. The procedure is the same for the two mixed qubits case, where no quantum correlated states are found. This is done by verifying the form of the state, i.e., if it can be written as a CC state it does not possess any quantum correlation, otherwise the correlations have some quantum nature. As shown in the main text, the synthesized algorithm is composed by: a Hadamard gate; one-qubit rotations around the $z$ axis, here indicated by $R_{k}^{i}$, where $k$ is an index associated to the rotation angle $\theta_{k}$ defined in Fig. \ref{fig:Synthesized-Deutsch-Jozsa-algori-1} and $i$ is the qubit index, beginning from $0$ for the semi-pure qubit and ranging from $1$ to $3$ for the mixed qubits; and controlled operations, here indicated by $CNOT_{m}^{n}$, where $m$ and $n$ are the indexes of the control and target qubit, respectively.

The initial state $\rho_{ini}=2^{-(n+1)}(I_{0}+\alpha Z_{0})\otimes I^{\otimes n}$ has no quantum correlations. Since $I_{0}$ can be written in any basis, including the eigenstates $Z_{0}$, this state can take the form $\rho_{ini}= \sum_i p_{i}\left|i\right\rangle \left\langle i\right|$, with $p_{i}=\{2^{-(n+1)}(1+\alpha),2^{-(n+1)}(1-\alpha)\}$ representing a classical probability distribution. 

Now we will present the states $\rho_{s}$ after each step $s$ of the synthesized algorithm (beginning from step $0$) and do a brief analysis of quantumness of correlations for some initial and final steps.

%\begin{widetext}

\thickmuskip=0mu

0) Hadamard gate on qubit $0$ and $R_{0}^{3}$: 
\[
\rho_{0}=2^{-4}(I_{0}+\alpha X_{0})\otimes I^{\otimes3},
\]
where we used $n=3$ to indicate that we are studying the three mixed
qubits case. This state can be put in the same form as the previous
one and also represents a classical probability distribution. Therefore, it does not have quantum correlations.

1) $CNOT_{0}^{3}$: 
\[
\rho_{1}=2^{-4}(I^{\otimes4}+\alpha X_{0}I_{1}I_{2}X_{3}),
\]
where we dropped some of the $\otimes$ for simplicity. Again, the
identities for the qubits 0 and 3 on the first term can be written
in the same basis of the $X_{0}$ and $X_{3}$ on the second term,
so the state can be written in a totally classical form.

2) $R_{1}^{3}$:

\[
\rho_{2}=2^{-4}(I^{\otimes4}+\alpha X_{0}I_{1}I_{2}Q_{3}(\theta_{1})),
\]
where every matrix $Q_{k}(x)$ has the form $\left|0\right\rangle \left\langle 1\right|_{k}e^{-ix}+\left|1\right\rangle \left\langle 0\right|_{k}e^{ix}$.
In this case, as the identity assumes the same form in any basis,
$I_{0}$ and $X_{0}$ are diagonal in the eigenbasis of $X_{0}$. Similarly $I_{3}$ and $Q_{3}(\theta_{1})$ have a common eigenbasis, so $\rho_{2}$
assumes a classical form.

3) $CNOT_{1}^{3}$:

\[
\rho_{3}=2^{-4}\left\{ I^{\otimes4}+\alpha X_{0}\left|0\right\rangle \left\langle 0\right|_{1}I_{2}Q_{3}(\theta_{1})+\alpha X_{0}\left|1\right\rangle \left\langle 1\right|_{1}I_{2}Q_{3}^{*}(\theta_{1})\right\} ,
\]
where $Q_{3}^{\dagger}(\theta_{1})$ is the Hermitian conjugate of $Q_{3}(\theta_{1})$.
The commutativity between $Q_{3}(\theta_{1})$ and $Q_{3}^{\dagger}(\theta_{1})$
depends on the value of $\theta_{1}$. From the main text we have that, for balanced functions, $\theta_{1}$ may assume a value of the set $\left\{ 0,\pm\nicefrac{\pi}{4},\pm\nicefrac{\pi}{2}\right\} $. If $\theta_{1}=\pm\nicefrac{\pi}{4}$, $Q_{3}(\theta_{1})$ and $Q_{3}^{\dagger}(\theta_{1})$ commute. Then, the state $\rho_{3}$ will be quantum correlated if $\theta_{1}=\pm\nicefrac{\pi}{4}$,
and will be a classical state if $\theta_{1}=0$ or $\theta_{1}=\pm\nicefrac{\pi}{2}$.
Thus, the system possesses quantum correlations at this point of
the synthesized algorithm only for balanced functions for which $\theta_{1}=\pm\nicefrac{\pi}{4}$.

4) $R_{2}^{3}$: 
\[
\rho_{4}=2^{-4}\left\{ I^{\otimes4}+\alpha X_{0}\left|0\right\rangle \left\langle 0\right|_{1}I_{2}Q_{3}(\theta_{2}+\theta_{1})+\alpha X_{0}\left|1\right\rangle \left\langle 1\right|_{1}I_{2}Q_{3}(\theta_{2}-\theta_{1})\right\} .
\]
As in the previous step, for some values of $\theta_{1}$ and $\theta_{2}$
(e.g. $\theta_{1}=\pm\nicefrac{\pi}{4}$ and $\theta_{2}=0$) this
state cannot be put in a diagonal form, such that, for some balanced
functions $\rho_{4}$ is quantum correlated.

5) $CNOT_{0}^{3}$: 
\[
\rho_{5}=2^{-4}\left\{ I^{\otimes4}+\left[\alpha\left|0\right\rangle \left\langle 1\right|_{0}\left[\left|0\right\rangle \left\langle 0\right|_{1}I_{2}P_{3}(\theta_{2}+\theta_{1})+\left|1\right\rangle \left\langle 1\right|_{1}I_{2}P_{3}(\theta_{2}-\theta_{1})\right]+H.c.\right]\right\} ,
\]
where $H.c.$ means the Hermitian conjugate and $P_{k}(x)=\left|0\right\rangle \left\langle 0\right|_{k}e^{-ix}+\left|1\right\rangle \left\langle 1\right|_{k}e^{ix}$.
Rearranging this expression we obtain $\rho_{5}=\left|0\right\rangle \left\langle 0\right|_{1}I_{2}[I_{0}I_{3} +\alpha\left|0\right\rangle \left\langle 1\right|_{0}P_{3}(\theta_{2}+\theta_{1})\\+\alpha\left|1\right\rangle \left\langle 0\right|_{0}P_{3}^{\dagger}(\theta_{2}+\theta_{1})]+\left|1\right\rangle \left\langle 1\right|_{1}I_{2}[I_{0}I_{3}+\alpha\left|0\right\rangle \left\langle 1\right|_{0}P_{3}(\theta_{2}-\theta_{1})+\alpha\left|1\right\rangle \left\langle 0\right|_{0}P_{3}^{\dagger}(\theta_{2}-\theta_{1})]$.
In all terms of $\rho_{5}$ the states for qubits $1$ and $2$ are
diagonal on the same basis (for each qubit state space), but the terms
for qubits $0$ and $3$ are not. The commutation between the two
terms on the right hand side of the previous expression is proportional
to $\left(\left|0\right\rangle \left\langle 0\right|_{0}-\left|1\right\rangle \left\langle 1\right|_{0}\right)[P_{3}(\theta_{2}+\theta_{1})P_{3}^{\dagger}(\theta_{2}-\theta_{1})-P_{3}(\theta_{2}-\theta_{1})P_{3}^{\dagger}(\theta_{2}+\theta_{1})]$,
which in turn is proportional to $\sin(2\theta_{1})$. Thus, if $\theta_{1}=\pm\nicefrac{\pi}{4}$
$\rho_{5}$ is quantum correlated at this point of the synthesized
algorithm, otherwise it represents just a classical probability
distribution. As the procedure to identify quantumness of correlations
is the same for the remaining states, we will just write down these
states and inform here, in advance, that all states up to $\rho_{28}$
can be quantum correlated for some balanced functions. We will show the analysis for the final states of the computation.

6) $R_{3}^{3}$:
\[
\rho_{6}=\rho_{5}.
\]

7) $CNOT_{0}^{3}$:
\begingroup
\setlength{\thickmuskip}{0mu}
\setlength{\medmuskip}{0mu}
\setlength{\thinmuskip}{0mu}
\begin{eqnarray*}
\rho_{7} & = & 2^{-4}\left\{ I^{\otimes4}+\left[\alpha\left|0\right\rangle \left\langle 1\right|_{0}\left[\left|0\right\rangle \left\langle 0\right|_{1}\left[\left|0\right\rangle \left\langle 0\right|_{2}P_{3}(\theta_{2}+\theta_{1})+\left|1\right\rangle \left\langle 1\right|_{2}P_{3}^{\dagger}(\theta_{2}+\theta_{1})\right]\right.\right.\right.\\
&  & \left.\left.\left.+\left|1\right\rangle \left\langle 1\right|_{1}\left[\left|0\right\rangle \left\langle 0\right|_{2}P_{3}(\theta_{2}-\theta_{1})+\left|1\right\rangle \left\langle 1\right|_{2}P_{3}^{\dagger}(\theta_{2}-\theta_{1})\right]\right]+H.c.\right]\right\} .
\end{eqnarray*}
\endgroup

8) $R_{4}^{3}$: 
\[
\rho_{8}=\rho_{7}.
\]

9) $CNOT_{1}^{3}$:
\begingroup
\setlength{\thickmuskip}{0mu}
\setlength{\medmuskip}{0mu}
\setlength{\thinmuskip}{0mu}
\begin{eqnarray*}
\rho_{9} & = & 2^{-4}\left\{ I^{\otimes4}+\left[\alpha\left|0\right\rangle \left\langle 1\right|_{0}\left[\left|0\right\rangle \left\langle 0\right|_{1}\left[\left|0\right\rangle \left\langle 0\right|_{2}P_{3}(\theta_{2}+\theta_{1})+\left|1\right\rangle \left\langle 1\right|_{2}P_{3}^{\dagger}(\theta_{2}+\theta_{1})\right]\right.\right.\right.\\
 &  & \left.\left.\left.+\left|1\right\rangle \left\langle 1\right|_{1}\left[\left|0\right\rangle \left\langle 0\right|_{2}P_{3}^{\dagger}(\theta_{2}-\theta_{1})+\left|1\right\rangle \left\langle 1\right|_{2}P_{3}(\theta_{2}-\theta_{1})\right]\right]+H.c.\right]\right\} .
\end{eqnarray*}
\endgroup

10) $R_{5}^{3}$: 
\[
\rho_{10}=\rho_{9}.
\]

11) $CNOT_{0}^{3}$:
\begingroup
\setlength{\thickmuskip}{0mu}
\setlength{\medmuskip}{0mu}
\setlength{\thinmuskip}{0mu}
\begin{eqnarray*}
\rho_{11} & = & 2^{-4}\left\{ I^{\otimes4}+\left[\alpha\left|0\right\rangle \left\langle 1\right|_{0}\left[\left|0\right\rangle \left\langle 0\right|_{1}\left[\left|0\right\rangle \left\langle 0\right|_{2}Q_{3}(\theta_{2}+\theta_{1})+\left|1\right\rangle \left\langle 1\right|_{2}Q_{3}^{\dagger}(\theta_{2}+\theta_{1})\right]\right.\right.\right.\\
 &  & \left.\left.\left.+\left|1\right\rangle \left\langle 1\right|_{1}\left[\left|0\right\rangle \left\langle 0\right|_{2}Q_{3}^{\dagger}(\theta_{2}-\theta_{1})+\left|1\right\rangle \left\langle 1\right|_{2}Q_{3}(\theta_{2}-\theta_{1})\right]\right]+H.c.\right]\right\} .
\end{eqnarray*}
\endgroup

12) $R_{6}^{3}$:
\begingroup
\setlength{\thickmuskip}{0mu}
\setlength{\medmuskip}{0mu}
\setlength{\thinmuskip}{0mu}
\begin{eqnarray*}
\rho_{12} & = & 2^{-4}\left\{ I^{\otimes4} \right. \\
& & \left.+\left[\alpha\left|0\right\rangle \left\langle 1\right|_{0}\left[\left|0\right\rangle \left\langle 0\right|_{1}\left[\left|0\right\rangle \left\langle 0\right|_{2}Q_{3}(\theta_{6}+\theta_{2}+\theta_{1}) + \left|1\right\rangle \left\langle 1\right|_{2}Q_{3}(\theta_{6}-\theta_{2}-\theta_{1})\right]\right.\right.\right.\\
 &  & \left.\left.\left.+\left|1\right\rangle \left\langle 1\right|_{1}\left[\left|0\right\rangle \left\langle 0\right|_{2}Q_{3}(\theta_{6}-\theta_{2}+\theta_{1})+\left|1\right\rangle \left\langle 1\right|_{2}Q_{3}(\theta_{6}+\theta_{2}-\theta_{1})\right]\right]+H.c.\right]\right\}.
\end{eqnarray*}
\endgroup

13) $CNOT_{1}^{3}$: 
\begingroup
\setlength{\thickmuskip}{0mu}
\setlength{\medmuskip}{0mu}
\setlength{\thinmuskip}{0mu}
\begin{eqnarray*}
\rho_{13} & = & 2^{-4}\left\{ I^{\otimes4} \right. \\
& & \left.+\left[\alpha\left|0\right\rangle \left\langle 1\right|_{0}\left[\left|0\right\rangle \left\langle 0\right|_{1}\left[\left|0\right\rangle \left\langle 0\right|_{2}Q_{3}(\theta_{6}+\theta_{2}+\theta_{1})+\left|1\right\rangle \left\langle 1\right|_{2}Q_{3}(\theta_{6}-\theta_{2}-\theta_{1})\right]\right.\right.\right.\\
 &  & \left.\left.\left.+\left|1\right\rangle \left\langle 1\right|_{1}\left[\left|0\right\rangle \left\langle 0\right|_{2}Q_{3}^{\dagger}(\theta_{6}-\theta_{2}+\theta_{1})+\left|1\right\rangle \left\langle 1\right|_{2}Q_{3}^{\dagger}(\theta_{6}+\theta_{2}-\theta_{1})\right]\right]+H.c.\right]\right\}.
\end{eqnarray*}
\endgroup

14) $R_{7}^{3}$: 
\begingroup
\setlength{\thickmuskip}{0mu}
\setlength{\medmuskip}{0mu}
\setlength{\thinmuskip}{0mu}
\begin{eqnarray*}
\rho_{14} & = & 2^{-4}\left\{ I^{\otimes4} \right. \\
& & \left.+\left[\alpha\left|0\right\rangle \left\langle 1\right|_{0}\left[\left|0\right\rangle \left\langle 0\right|_{1}\left[\left|0\right\rangle \left\langle 0\right|_{2}Q_{3}(\theta_{7}+\theta_{6}+\theta_{2}+\theta_{1}) +\left|1\right\rangle \left\langle 1\right|_{2}Q_{3}(\theta_{7}+\theta_{6}-\theta_{2}-\theta_{1})\right]\right.\right.\right.\\
 &  & \left.\left.\left.+\left|1\right\rangle \left\langle 1\right|_{1}\left[\left|0\right\rangle \left\langle 0\right|_{2}Q_{3}(\theta_{7}-\theta_{6}+\theta_{2}-\theta_{1})+\left|1\right\rangle \left\langle 1\right|_{2}Q_{3}(\theta_{7}-\theta_{6}-\theta_{2}+\theta_{1})\right]\right]+H.c.\right]\right\}.
\end{eqnarray*}
\endgroup

15) $CNOT_{2}^{3}$:
\begingroup
\setlength{\thickmuskip}{0mu}
\setlength{\medmuskip}{0mu}
\setlength{\thinmuskip}{0mu}
\begin{eqnarray*}
\rho_{15} & = & 2^{-4}\left\{ I^{\otimes4} \right. \\
& & \left.+\left[\alpha\left|0\right\rangle \left\langle 1\right|_{0}\left[\left|0\right\rangle \left\langle 0\right|_{1}\left[\left|0\right\rangle \left\langle 0\right|_{2}Q_{3}(\theta_{7}+\theta_{6}+\theta_{2}+\theta_{1}) +\left|1\right\rangle \left\langle 1\right|_{2}Q_{3}^{\dagger}(\theta_{7}+\theta_{6}-\theta_{2}-\theta_{1})\right]\right.\right.\right.\\
 &  & \left.\left.\left.+\left|1\right\rangle \left\langle 1\right|_{1}\left[\left|0\right\rangle \left\langle 0\right|_{2}Q_{3}(\theta_{7}-\theta_{6}+\theta_{2}-\theta_{1})+\left|1\right\rangle \left\langle 1\right|_{2}Q_{3}^{\dagger}(\theta_{7}-\theta_{6}-\theta_{2}+\theta_{1})\right]\right]+H.c.\right]\right\}.
\end{eqnarray*}
\endgroup

16) $CNOT_{1}^{3}$:
\begingroup
\setlength{\thickmuskip}{0mu}
\setlength{\medmuskip}{0mu}
\setlength{\thinmuskip}{0mu}
\begin{eqnarray*}
\rho_{16} & = & 2^{-4}\left\{ I^{\otimes4} \right. \\
& & \left.+\left[\alpha\left|0\right\rangle \left\langle 1\right|_{0}\left[\left|0\right\rangle \left\langle 0\right|_{1}\left[\left|0\right\rangle \left\langle 0\right|_{2}Q_{3}(\theta_{7}+\theta_{6}+\theta_{2}+\theta_{1})+\left|1\right\rangle \left\langle 1\right|_{2}Q_{3}^{\dagger}(\theta_{7}+\theta_{6}-\theta_{2}-\theta_{1})\right]\right.\right.\right.\\
 &  & \left.\left.\left.+\left|1\right\rangle \left\langle 1\right|_{1}\left[\left|0\right\rangle \left\langle 0\right|_{2}Q_{3}^{\dagger}(\theta_{7}-\theta_{6}+\theta_{2}-\theta_{1})+\left|1\right\rangle \left\langle 1\right|_{2}Q_{3}(\theta_{7}-\theta_{6}-\theta_{2}+\theta_{1})\right]\right]+H.c.\right]\right\}.
\end{eqnarray*}
\endgroup

17) $CNOT_{2}^{3}$:
\begingroup
\setlength{\thickmuskip}{0mu}
\setlength{\medmuskip}{0mu}
\setlength{\thinmuskip}{0mu}
\begin{eqnarray*}
\rho_{17} & = & 2^{-4}\left\{ I^{\otimes4} \right. \\
& & \left.+\left[\alpha\left|0\right\rangle \left\langle 1\right|_{0}\left[\left|0\right\rangle \left\langle 0\right|_{1}\left[\left|0\right\rangle \left\langle 0\right|_{2}P_{3}(\theta_{7}+\theta_{6}+\theta_{2}+\theta_{1})+\left|1\right\rangle \left\langle 1\right|_{2}P_{3}^{\dagger}(\theta_{7}+\theta_{6}-\theta_{2}-\theta_{1})\right]\right.\right.\right.\\
 &  & \left.\left.\left.+\left|1\right\rangle \left\langle 1\right|_{1}\left[\left|0\right\rangle \left\langle 0\right|_{2}P_{3}^{\dagger}(\theta_{7}-\theta_{6}+\theta_{2}-\theta_{1})+\left|1\right\rangle \left\langle 1\right|_{2}P_{3}(\theta_{7}-\theta_{6}-\theta_{2}+\theta_{1})\right]\right]+H.c.\right]\right\} .
\end{eqnarray*}
\endgroup

Let us define now $A_{3} \equiv P_{3}(\theta_{7}+\theta_{6}+\theta_{2}+\theta_{1})$,
$B_{3} \equiv P_{3}^{\dagger}(\theta_{7}+\theta_{6}-\theta_{2}-\theta_{1})$, $C_{3} \equiv P_{3}^{\dagger}(\theta_{7}-\theta_{6}+\theta_{2}-\theta_{1})$
and $D_{3} \equiv P_{3}(\theta_{7}-\theta_{6}-\theta_{2}+\theta_{1})$.

18) $R_{8}^{2}$: 
\[
\rho_{18}=\rho_{17}.
\]

19) $CNOT_{0}^{2}$: 
\begingroup
\begin{center}
\setlength{\thickmuskip}{0mu}
\setlength{\medmuskip}{0mu}
\setlength{\thinmuskip}{0mu}
\begin{eqnarray*}
\rho_{19}&=&2^{-4}\left\{ I^{\otimes4}+\left[\alpha\left|0\right\rangle \left\langle 1\right|_{0}\left[\left|0\right\rangle \left\langle 0\right|_{1}\left[\left|0\right\rangle \left\langle 1\right|_{2}A_{3}+\left|1\right\rangle \left\langle 0\right|_{2}B_{3}\right]\right.\right.\right.\\
 &  & \left.\left.\left. +\left|1\right\rangle \left\langle 1\right|_{1}\left[\left|0\right\rangle \left\langle 1\right|_{2}C_{3}+\left|1\right\rangle \left\langle 0\right|_{2}D_{3}\right]\right]+H.c.\right]\right\} .
\end{eqnarray*}
\end{center}
\endgroup

20) $R_{9}^{2}$:
\begingroup
\setlength{\thickmuskip}{0mu}
\setlength{\medmuskip}{0mu}
\setlength{\thinmuskip}{0mu}
\begin{eqnarray*}
\rho_{20} & = & 2^{-4}\left\{ I^{\otimes4}+\left[\alpha\left|0\right\rangle \left\langle 1\right|_{0}\left[\left|0\right\rangle \left\langle 0\right|_{1}\left[e^{-i\theta_{9}}\left|0\right\rangle \left\langle 1\right|_{2}A_{3}+e^{i\theta_{9}}\left|1\right\rangle \left\langle 0\right|_{2}B_{3}\right]\right.\right.\right.\\
 &  & \left.\left.\left.+\left|1\right\rangle \left\langle 1\right|_{1}\left[e^{-i\theta_{9}}\left|0\right\rangle \left\langle 1\right|_{2}C_{3}+e^{i\theta_{9}}\left|1\right\rangle \left\langle 0\right|_{2}D_{3}\right]\right]+H.c.\right]\right\} .
\end{eqnarray*}
\endgroup

21) $CNOT_{1}^{2}$:
\begingroup
\setlength{\thickmuskip}{0mu}
\setlength{\medmuskip}{0mu}
\setlength{\thinmuskip}{0mu}
\begin{eqnarray*}
\rho_{21} & = & 2^{-4}\left\{ I^{\otimes4}+\left[\alpha\left|0\right\rangle \left\langle 1\right|_{0}\left[\left|0\right\rangle \left\langle 0\right|_{1}\left[e^{-i\theta_{9}}\left|0\right\rangle \left\langle 1\right|_{2}A_{3}+e^{i\theta_{9}}\left|1\right\rangle \left\langle 0\right|_{2}B_{3}\right]\right.\right.\right.\\
 &  & \left.\left.\left.+\left|1\right\rangle \left\langle 1\right|_{1}\left[e^{-i\theta_{9}}\left|1\right\rangle \left\langle 0\right|_{2}C_{3}+e^{i\theta_{9}}\left|0\right\rangle \left\langle 1\right|_{2}D_{3}\right]\right]+H.c.\right]\right\} .
\end{eqnarray*}
\endgroup

22) $R_{10}^{2}$:
\begingroup
\setlength{\thickmuskip}{0mu}
\setlength{\medmuskip}{0mu}
\setlength{\thinmuskip}{0mu}
\begin{eqnarray*}
\rho_{22} & = & 2^{-4}\left\{ I^{\otimes4}+\left[\alpha\left|0\right\rangle \left\langle 1\right|_{0}\left[\left|0\right\rangle \left\langle 0\right|_{1}\left[e^{-i\left(\theta_{10}+\theta_{9}\right)}\left|0\right\rangle \left\langle 1\right|_{2}A_{3}+e^{i\left(\theta_{10}+\theta_{9}\right)}\left|1\right\rangle \left\langle 0\right|_{2}B_{3}\right]\right.\right.\right.\\
 &  & \left.\left.\left.+\left|1\right\rangle \left\langle 1\right|_{1}\left[e^{i\left(\theta_{10}-\theta_{9}\right)}\left|1\right\rangle \left\langle 0\right|_{2}C_{3}+e^{-i\left(\theta_{10}-\theta_{9}\right)}\left|0\right\rangle \left\langle 1\right|_{2}D_{3}\right]\right]+H.c.\right]\right\} .
\end{eqnarray*}
\endgroup

23) $CNOT_{0}^{2}$:
\begingroup
\setlength{\thickmuskip}{0mu}
\setlength{\medmuskip}{0mu}
\setlength{\thinmuskip}{0mu}
\begin{eqnarray*}
\rho_{23} & = & 2^{-4}\left\{ I^{\otimes4}+\left[\alpha\left|0\right\rangle \left\langle 1\right|_{0}\left[\left|0\right\rangle \left\langle 0\right|_{1}\left[e^{-i\left(\theta_{10}+\theta_{9}\right)}\left|0\right\rangle \left\langle 0\right|_{2}A_{3}+e^{i\left(\theta_{10}+\theta_{9}\right)}\left|1\right\rangle \left\langle 1\right|_{2}B_{3}\right]\right.\right.\right.\\
 &  & \left.\left.\left.+\left|1\right\rangle \left\langle 1\right|_{1}\left[e^{i\left(\theta_{10}-\theta_{9}\right)}\left|1\right\rangle \left\langle 1\right|_{2}C_{3}+e^{-i\left(\theta_{10}-\theta_{9}\right)}\left|0\right\rangle \left\langle 0\right|_{2}D_{3}\right]\right]+H.c.\right]\right\} .
\end{eqnarray*}
\endgroup

24) $R_{11}^{2}$:
\[
\rho_{24}=\rho_{23}.
\]

25) $CNOT_{1}^{2}$:
\begingroup
\setlength{\thickmuskip}{0mu}
\setlength{\medmuskip}{0mu}
\setlength{\thinmuskip}{0mu}
\begin{eqnarray*}
\rho_{25} & = & 2^{-4}\left\{ I^{\otimes4}+\left[\alpha\left|0\right\rangle \left\langle 1\right|_{0}\left[\left|0\right\rangle \left\langle 0\right|_{1}\left[e^{-i\left(\theta_{10}+\theta_{9}\right)}\left|0\right\rangle \left\langle 0\right|_{2}A_{3}+e^{i\left(\theta_{10}+\theta_{9}\right)}\left|1\right\rangle \left\langle 1\right|_{2}B_{3}\right]\right.\right.\right.\\
 &  & \left.\left.\left.+\left|1\right\rangle \left\langle 1\right|_{1}\left[e^{i\left(\theta_{10}-\theta_{9}\right)}\left|0\right\rangle \left\langle 0\right|_{2}C_{3}+e^{-i\left(\theta_{10}-\theta_{9}\right)}\left|1\right\rangle \left\langle 1\right|_{2}D_{3}\right]\right]+H.c.\right]\right\} .
\end{eqnarray*}
\endgroup

26) $R_{12}^{1}$:
\[
\rho_{26}=\rho_{25}.
\]

27) $CNOT_{0}^{1}$:
\begingroup
\setlength{\thickmuskip}{0mu}
\setlength{\medmuskip}{0mu}
\setlength{\thinmuskip}{0mu}
\begin{eqnarray*}
\rho_{27} & = & 2^{-4}\left\{ I^{\otimes4}+\left[\alpha\left|0\right\rangle \left\langle 1\right|_{0}\left[\left|0\right\rangle \left\langle 1\right|_{1}\left[e^{-i\left(\theta_{10}+\theta_{9}\right)}\left|0\right\rangle \left\langle 0\right|_{2}A_{3}+e^{i\left(\theta_{10}+\theta_{9}\right)}\left|1\right\rangle \left\langle 1\right|_{2}B_{3}\right]\right.\right.\right.\\
 &  & \left.\left.\left.+\left|1\right\rangle \left\langle 0\right|_{1}\left[e^{i\left(\theta_{10}-\theta_{9}\right)}\left|0\right\rangle \left\langle 0\right|_{2}C_{3}+e^{-i\left(\theta_{10}-\theta_{9}\right)}\left|1\right\rangle \left\langle 1\right|_{2}D_{3}\right]\right]+H.c.\right]\right\} .
\end{eqnarray*}
\endgroup

28) $R_{13}^{1}$:
\begingroup
\setlength{\thickmuskip}{0mu}
\setlength{\medmuskip}{0mu}
\setlength{\thinmuskip}{0mu}
\begin{eqnarray*}
\rho_{28} & = & 2^{-4}\left\{ I^{\otimes4} \right. \\
& & \left.+\left[\alpha\left|0\right\rangle \left\langle 1\right|_{0}\left[e^{-i\theta_{13}}\left|0\right\rangle \left\langle 1\right|_{1}\left[e^{-i\left(\theta_{10}+\theta_{9}\right)}\left|0\right\rangle \left\langle 0\right|_{2}A_{3}+e^{i\left(\theta_{10}+\theta_{9}\right)}\left|1\right\rangle \left\langle 1\right|_{2}B_{3}\right]\right.\right.\right.\\
 &  & \left.\left.\left.+e^{i\theta_{13}}\left|1\right\rangle \left\langle 0\right|_{1}\left[e^{i\left(\theta_{10}-\theta_{9}\right)}\left|0\right\rangle \left\langle 0\right|_{2}C_{3}+e^{-i\left(\theta_{10}-\theta_{9}\right)}\left|1\right\rangle \left\langle 1\right|_{2}D_{3}\right]\right]+H.c.\right]\right\} .
\end{eqnarray*}
\endgroup
This state is clearly diagonal on qubits $2$ and $3$, but it is not fully diagonal on qubits $0$ and $1$ for balanced functions.
Thus, the correlations in state $\rho_{28}$ may present some quantum nature.

29) $CNOT_{0}^{1}$:
\begingroup
\setlength{\thickmuskip}{0mu}
\setlength{\medmuskip}{0mu}
\setlength{\thinmuskip}{0mu}
\begin{eqnarray*}
\rho_{29} & = & 2^{-4}\left\{ I^{\otimes4} \right. \\
& & \left.+\left[\alpha\left|0\right\rangle \left\langle 1\right|_{0}\left[e^{-i\theta_{13}}\left|0\right\rangle \left\langle 0\right|_{1}\left[e^{-i\left(\theta_{10}+\theta_{9}\right)}\left|0\right\rangle \left\langle 0\right|_{2}A_{3}+e^{i\left(\theta_{10}+\theta_{9}\right)}\left|1\right\rangle \left\langle 1\right|_{2}B_{3}\right]\right.\right.\right.\\
 &  & \left.\left.\left.+e^{i\theta_{13}}\left|1\right\rangle \left\langle 1\right|_{1}\left[e^{i\left(\theta_{10}-\theta_{9}\right)}\left|0\right\rangle \left\langle 0\right|_{2}C_{3}+e^{-i\left(\theta_{10}-\theta_{9}\right)}\left|1\right\rangle \left\langle 1\right|_{2}D_{3}\right]\right]+H.c.\right]\right\} .
\end{eqnarray*}
\endgroup
The $CNOT_{0}^{1}$ turns the state diagonal on qubit $1$, besides
being already diagonal on qubits $2$ and $3$. Now, the state of qubit $0$ in each term of the expression for $\rho_{29}$ admits
the same eigenbasis, so the correlations in the total state are purely classical
for any balanced or constant function.

30) $e^{i\Phi}R_{14}^{0}$:
\begingroup
\setlength{\thickmuskip}{0mu}
\setlength{\medmuskip}{0mu}
\setlength{\thinmuskip}{0mu}
\begin{eqnarray*}
\rho_{30} & = & 2^{-4}\left\{ I^{\otimes4} \right. \\
& & \left.+\left[\alpha e^{-i\theta_{14}}\left|0\right\rangle \left\langle 1\right|_{0}\left[e^{-i\theta_{13}}\left|0\right\rangle \left\langle 0\right|_{1}\left[e^{-i\left(\theta_{10}+\theta_{9}\right)}\left|0\right\rangle \left\langle 0\right|_{2}A_{3}+e^{i\left(\theta_{10}+\theta_{9}\right)}\left|1\right\rangle \left\langle 1\right|_{2}B_{3}\right]\right.\right.\right.\\
 &  & \left.\left.\left.+e^{i\theta_{13}}\left|1\right\rangle \left\langle 1\right|_{1}\left[e^{i\left(\theta_{10}-\theta_{9}\right)}\left|0\right\rangle \left\langle 0\right|_{2}C_{3}+e^{-i\left(\theta_{10}-\theta_{9}\right)}\left|1\right\rangle \left\langle 1\right|_{2}D_{3}\right]\right]+H.c.\right]\right\} .
\end{eqnarray*}
\endgroup
The application of this last gate does not generate any quantum correlation
in the final state of the system as is shown in the main text.

\end{document}